\documentclass[12pt]{article}
\usepackage{epsfig}
\begin{document}
\begin{center}
{\Large {\bf Gauge invariant discretization of Poincare quantum gravity \\

}} \vskip-40mm \rightline{\small ITEP-LAT 2006-03} \vskip 30mm

{
\vspace{1cm}
{ M.A.~Zubkov$^a$  }\\

 \vspace{.5cm} { \it
$^a$ ITEP, B.Cheremushkinskaya 25, Moscow, 117259, Russia }}
\end{center}
\begin{abstract}
In this paper we suggest gauge invariant discretization of Poincare quantum
gravity. We generalize  Regge calculus to the case of Riemann-Cartan space. The
basic element of the constructed discretization is piecewize linear
Riemann-Cartan space with flat pieces of hypercubic form. We consider the model
with squared curvature action and calculate the correspondent lattice action.
We construct local  measure over the dynamical variables of the lattice model.
\end{abstract}

\today \hspace{1ex}

\newpage

\section{Introduction}

In order to put quantum gravity theory on the lattice one should construct such
discretization of the continuum model, which preserves as much symmetry of the
original continuum model as possible. That's why Regge quantum gravity is
considered  to be one of the most natural lattice realizations of quantum
gravity. The main success of Regge calculus is that it is manifestly gauge
invariant, i.e. it preserves the whole symmetry of the original model. This is
due to the fact, that in this lattice theory piecewize - linear Riemannian
manyfold plays the role of the dynamical variable. (In the original continuum
theory Riemannian manyfold of any possible form plays the role of the dynamical
variable.) In other words, we approximate any given Riemannian manifold by
piecewize - linear Riemannian manyfolds. This discretization is gauge invariant
by the construction.

In order to define measure over link lengths in Regge quantum gravity one may
start from the invariant metric on the space of continuum Riemannian geometries
\cite{DeWitt,Misner,Popov}:
\begin{equation}
\|\delta g\|^2 = \frac{1}{2}\int d^D x \sqrt{|g|} (g^{\mu \nu}g^{\rho \eta} +
g^{\rho \nu}g^{\mu \eta} + C g^{\mu \rho}g^{\nu \eta})\delta g_{\mu \rho}
\delta g_{\nu \eta}\label{NORM}
\end{equation}
where $C$ is the arbitrary constant such that $C\ne-2/D$. Then we may postulate
that the correct measure on the space of Riemannian geometries corresponds to
this metric in the same sense as the measure $dx$ on $R^1$ corresponds to the
metric $\|\delta x\|^2 = (\delta x)^2$. Unfortunately, the shift from finite
dimensional case to infinite dimensional case is not clear. Namely, we may
generalize the finite dimensional formula for the measure correspondent to the
given metric. Thus we obtain the continuum measure correspondent to the metric
(\ref{NORM}) (for more details see, for example, \cite{W} and references
therein):
\begin{equation}
Dg  =  \Pi_x (\sqrt{|g(x)|})^{\sigma} \Pi_{\mu \ge \nu}
dg_{\mu\nu}(x)\label{Measure}
\end{equation}
Here the choice $\sigma = (D-4)(D+1)/4$ corresponds to the super-metric tensor
in its form chosen in \cite{DeWitt,Fijikawa} while the choice $\sigma=-(D+1)$
was considered in \cite{Misner}. Obviously this ambiguity is related to the
fact, that the product over points $\Pi_x$ is not well defined if $x$ belongs
to $R^D$. Actually, the precise meaning could be given to (\ref{Measure}) only
when the discretization of Riemannian space is chosen.

In some of the papers devoted to Regge quantum gravity, the following measure
over link lengths was considered (see, for example, \cite{W}):
\begin{equation}
Dl  =  [\Pi_x V_x^{\sigma}] [\Pi_{ij} dl^2_{ij}]
\Theta(l_{ij})\label{MeasureL},
\end{equation}
where the product $\Pi_x$ is over the simplices while the product $\Pi_{ij}$ is
over the vertices of the simplicial manyfold. $\Theta(l_{ij})$ is the step
function, which provides the triangle inequalities at each simplex. $V_x$ is
the volume of the simplex $x$, and $l_{ij}$ is the length of the link that
connects vertices $i$ and $j$.

According to \cite{W} the derivation  of the measure (\ref{MeasureL}) is as
follows. At each simplex $V_x^{\sigma} \Pi_{ij} dl^2_{ij}(x)$  is the direct
discretization of the expression \\
$(\sqrt{|g(x)|})^{\sigma} \Pi_{\mu \ge \nu} dg_{\mu\nu}(x)$ (where $i$ and $j$
are the vertices of the given simplex, and the constant factor is omitted). In
the system of simplices glued together the link lengths are not independent.
Therefore, the correspondent constraint should be imposed. In \cite{W} this
constraint was chosen in the form: $\Pi_{xy} \Pi_{ij}
\delta(l_{ij}^2(x)-l_{ij}^2(y))$, where the product $\Pi_{xy}$ is over the
pairs of neighboring simplices while the product $\Pi_{ij}$ is over the pairs
of vertices of the side, which is common for $x$ and $y$. Thus we obtain the
lattice measure in the form (\ref{MeasureL}).

We mention here again, that the derivation of the continuum formula
(\ref{Measure}) and even its own form cannot be considered as completely
mathematically rigorous.  There is also an ambiguity in the choice of the
constraint, which is imposed on the link lengths in the system of simplices
glued together. Namely, such a constraint may be choosen in the form
$\Omega[l_{ij}]\Pi_{xy} \Pi_{ij} \delta(l_{ij}^2(x)-l_{ij}^2(y))$, where
$\Omega[l_{ij}]$ depends upon link lengths. A priory the particular form of
such functional is not known.  Therefore, although the choice (\ref{MeasureL})
seems to be quite natural, it is not derived rigorously from the expression
(\ref{NORM}) for the metric on space of continuum Riemannian geometries. This
results, in particular, in the ambiguity in the choice of $\sigma$.

A possible solution for this problem would appear if there exists a symmetry in
the continuum theory, which (after transferring to discretized model) fixes the
form of lattice measure. Unfortunately, the existence of such a symmetry within
Regge quantum gravity is not clear. Therefore, in the present paper we suggest
to consider its generalization based on the so-called Poincare quantum gravity.
We hope that the mentioned problem can be solved within this generalized model.
This is the theory, in which the dynamical variable is the Riemann - Cartan
manyfold. (Riemannian manyfold is the limiting case of Riemann - Cartan
manyfold with vanishing torsion.)

It is worth mentioning, that the topics related to the discretization of
Poincare gravity were already considered. Say, in \cite{Magnea,Drummond} the
independent  $SO(4)$ connection was introduced into the Regge calculus. The
connection is singular and lives  on the sides of the simplices. In principle
this construction resembles the one suggested in the present paper. However,
here we use hypercubic lattice, which is much more useful, than the simplicial
one. Moreover, in the present paper we construct simple natural Poincare
invariant measure over the dynamical variables, which was not done in
\cite{Magnea,Drummond}. We also mention, that in \cite{Magnea,Drummond} there
was no consideration of the squared curvature terms in the action.

Poincare gravity on the hypercubic lattice was considered, say, in
\cite{Tomboulis, Magnea, Smolin, Manion, Kaku, Menotti}. In these papers
lattice discretization was constructed in the conventional way via direct
discretization of the theory. Therefore, the correspondent constructions were
not gauge invariant (in the gravity models the gauge group is the group of
general coordinate transformations)\footnote{It was shown, for example, that in
ordinary gauge theories the discretization, which is not gauge invariant, is
not appropriate (say, in the correspondent lattice nonabelian gauge models
there is no confinement of fundamental charges \cite{NONABELIAN}).}. In the
mentioned papers some of the considered lattice theories were invariant under
lattice gauge transformations, which belong to the Poincare group. However, the
existence of this invariance does not mean that the constructed discretizations
are invariant under general coordinate transformations. For a brief comparison
of our approach  with the approaches of \cite{Tomboulis, Magnea, Smolin,
Manion, Kaku, Menotti} see section $3$ of the present paper.

 In the present paper we generalize Regge construction to
Riemann-Cartan spaces. In our approach we approximate any given manyfold by a
piecewize-linear Riemann-Cartan manyfold (with hypercubic elements). The
crucial difference between our approach and the approachs of \cite{Tomboulis,
Magnea, Smolin, Manion, Kaku, Menotti} is that we use geometrical construction.
I.e. we discretize the original manyfold by the invariant  objects (with
respect to the group of general coordinate transformations). As a result our
lattice model is gauge invariant by construction.

\section{The discretization of Riemann-Cartan space}
Thus, we consider  Riemann-Cartan space, in which both $SO(4)$
 connection\footnote{Further we imply, that Wick rotation to Euclidean
 signature is performed. So, we deal with Euclidean path integral formalism.}
  and the inverse vierbein are the dynamical variables. We do not require vanishing of torsion (which would
lead to appearance of Riemannian manyfold) or vanishing of curvature (which
would lead to appearance of Weitzenbock space).    The discretized space is in
itself a Riemann-Cartan space.  It is composed of flat pieces connected
together. We consider the case, when each such piece has the hypercubic form.
 Further  we refer to hypercubes as to elements of the lattice.
  Form of the lattice elements is fixed by the set of vectors ${\bf e}_{\mu}$ that connect
the center of the element with its vertices. The expression of ${\bf e}_{\mu}$
through elements of the orthonormal frame ${\bf f}_A$ ($A = 1,2,3,4$) (common
for all lattice elements) is one of the basic variables of the construction. So
we have
\begin{equation}
{\bf e}_{\mu} = \sum_A E^A_{\mu} {\bf f}_A
\end{equation}
(Everywhere space - time indices are denoted by Greek letters contrary to the
tetrad ones.) We imply here that  vectors ${\bf e}_{\mu} (\mu = 0, 1,2,3,4)$
are independent.  The other vectors ${\bf e}_{\mu} (\mu = 5, ...,15)$ are
defined in such a way, that opposite sides of the lattice element are parallel
to each other. The hypercubic lattice is periodic and the position of the
starting point of each lattice element is always denoted by ${\bf e}_0$.
Vectors ${\bf e}_{\mu} (\mu =  1,2,3,4)$ point to its neighbors.

Variables $E^A_{\mu}$ represent  translations from the center of the lattice
element to its vertices. Metric (or vierbein) is implied to be constant inside
each lattice element. Here shift of the center of lattice element by a vector
$v^A$ causes transformation of basic variables: $E^A_{\mu} \rightarrow
E^A_{\mu} + v^A$, which could be treated as gauge transformation with respect
to the translational gauge group. It represents the translation of the given
lattice element within the correspondent local map.

In addition to the translational connection, which is defined by the set of
variables $E^A_{\mu}$, each shift from one lattice element to another is
accompanied by the rotation in the four - dimensional tangent space. In other
words, there is the $SO(4)$ connection, which is singular and is concentrated
on the sides of lattice elements\footnote{In this paper we do not consider
parity transformations, correspondent to the matrices ${\rm diag}\, \{\pm 1,\pm
1,\pm 1,\pm 1\}$ with negative determinant since these transformations do not
appear in the infinitesimal form of Poincare transformation. However, the
considered model could be easily generalized in order to include parity
transformations if we use $O(4)$ matrices instead of $SO(4)$ ones.}. We denote
by $U_{IJ}$ the $SO(4)$ matrix, which is attached to the side that is common
for the lattice elements $I$ and $J$.

The constructed Riemann - Cartan space has singular connection. In this case
definitions of curvature and torsion become ambiguous. Therefore we must fix
one of the definitions. For the details we refer to the Appendix, where one of
the definitions is used in order to calculate torsion and curvature on the
piecewize-linear manyfold.

The gauge transformations with respect to the whole Poincare group are
represented by translations and $SO(4)$ rotations of the lattice elements, that
result in the following change of basic variables:
\begin{equation}
E^A_{\mu} \rightarrow \Theta^A_B E^B_{\mu} + v^A,\label{trans}
\end{equation}
where $\Theta$ is the rotation matrix and $v$ is the vector that represents
translation.

\section{The action}

In this paper we consider the following action:

\begin{eqnarray}
S &=& \int \{ \alpha (R_{AB} R_{AB} - \frac{1}{3}R^2) + \beta R^2 - \gamma
m_p^2 R + \lambda m_P^4\} |{\cal E}| d^4 x \nonumber\\&& +\,  \delta m_P^2 \int
T^A_{BC} T^A_{BC} |{\cal E}| d^4 x, \label{S2}
\end{eqnarray}
where $|{\cal E}| = {\rm det} {\cal E}^A_{\mu}$, ${\cal E}^A_{\mu}$ is the
inverse vierbein, the tetrad components of Ricci tensor are denoted by
$R_{AB}$, and $R$ is the scalar curvature. Coupling constants $\alpha, \beta,
\gamma$ and $\lambda$ are dimensionless while $m_p$ is a dimensional parameter.
The second term is added in order to suppress torsion at $\delta
\rightarrow\infty$. So, at $\delta \rightarrow\infty$ we arrive at the model
considered in \cite{renormalizable}. This model is renormalizable and
asymptotic free at certain values of the coupling constants. Thus the model
with the action (\ref{S2}) is worth considering. More arguments in favor of
this point of view could be found in \cite{TELEPARALLEL}.

The action (\ref{S2}) can be calculated on the piecewize - linear
Riemann-Cartan manyfold composed of flat hypercubic cells. Below we represent
the resulting expression. For the details of the definition and calculation of
singular curvature and torsion on this manyfold see Appendix.

In order to rewrite in a useful form the expressions (\ref{RB}) and
(\ref{TSING}) for lattice curvature and toirsion we drop to the dual lattice.
Then our rotation matrices are attached to links while the inverse vierbein is
attached to sites. Let us denote by $U_{\mu}(x)$ the matrix correspondent to
the link, which begins at the site $x$ and points to the direction $\mu$ ($\mu
= \pm 4, \pm 3, \pm 2, \pm1$).   We denote by $\Omega_{\mu \nu}(x) = U_{\mu}(x)
... $ the product of link matrices along the boundary of the plaquette, which
is placed in the $(\mu \nu)$ plane. The inverse vierbein, which is attached to
the site $x$, is ${\cal E}^A_{\mu} = E^A_{\mu} - E^A_0, \mu = 1,2,3,4$. For
negative values of $\mu$ we define ${\cal E}^A_{-\mu} = - {\cal E}^A_{\mu}$.
The inverse matrix for positive values of indices is denoted by ${\cal
E}^{\mu}_A(x) = \{{\cal E}(x)^{-1}\}^{\mu}_A$. We also expand this definition
to negative values of indices: ${\cal E}^{\mu}_A(x) = {\rm sign}(\mu)\{[{\cal
E}(x)]^{-1}\}^{|\mu|}_A$. We shall denote by $\Delta x_{\mu}$ the shift on the
lattice by one step in the $\mu$ -th direction ($\Delta x_{-\mu} = - \Delta
x_{\mu} $). So, $x+\Delta x_{\mu}$ is the site which is obtained via the shift
from the site $x$ by one lattice spacing in the direction $\mu$ while $x-\Delta
x_{\mu}$ is obtained by the shift in the opposite direction. Thus, $U_{-\mu}(x)
= U^{-1}_{\mu}(x-\Delta x_{\mu})$. Next, we define $\Delta_{\mu} {\cal
E}_{\nu}(x) = U_{\mu}(x){\cal E}_{\nu}(x+\Delta x_{\mu}) - {\cal E}_{\nu}(x)$.
Everywhere we imply summation over the repeated indices. The summation over
space - time indices $\mu,\nu,...$ is implied over $\pm 4, \pm 3, \pm 2, \pm1$.

We introduce lattice tetrad components of torsion and curvature attached to the
sites of the lattice and to positive or negative directions (in this paragraph
and in the next paragraph the summation over Greek indices is not implied):
\begin{eqnarray}
&& {\bf [R^{\mu\nu}]}_{C F B}^A(x) = \frac{1}{2}{\cal E}^{\mu}_C {\cal E}^{\nu}_F [\Omega_{\mu \nu}-\Omega_{\nu \mu}]^A_B \nonumber\\
&& {\bf [R^{\mu\nu}]}_{F B}(x) = {\bf [R^{\mu\nu}]}_{A F B}^A (x)\nonumber\\
&& {\bf [R^{\mu\nu}]}(x) = {\bf [R^{\mu\nu}]}_{A A} (x)\nonumber\\
&& {\bf R}(x) = \sum_{\mu,\nu} {\bf [R^{\mu\nu}]} (x)\nonumber\\
&& {\bf [T^{\mu}]}^A_{C F}(x) = \sum_{\nu}\Delta_{\mu} {\cal E}^A_{\nu} [{\cal
E}^{\mu}_C {\cal E}^{\nu}_F-{\cal E}^{\nu}_C {\cal E}^{\mu}_F]
\end{eqnarray}

Now we are ready to express the action (\ref{S2}) on the piecewise-linear
manyfold through its parameters $E^A_{\mu}(x)$ and $U_{\mu}(x)$. In order to do
this we use the expressions for torsion and curvature (\ref{RB}) and
(\ref{TSING}). We omit the intermediate steps and represent the final result
using the notations for lattice torsion and curvature introduced above.
\begin{eqnarray}
S &=& \sum_{x\in sites} |{\cal E}(x)| {\large \bf ( } \sum_{\mu,\nu} \{
\bar{\alpha} {\bf [R^{\mu\nu}]}_{FB}(x){\bf [R^{\mu\nu}]}_{FB}(x) +
(\bar{\beta}-\frac{1}{3}\bar{\alpha}){\bf [R^{\mu\nu}]^2}(x)\nonumber\\&&
 - \bar{\gamma} m_p^2 {\bf [R^{\mu\nu}]}(x) \}  +\, \bar{\delta} m_P^2 \sum_{\mu}{\bf
[T^{\mu}]}^A_{BC}(x){\bf [T^{\mu}]}^A_{BC}(x)+ \bar{\lambda} m_P^4 {\large \bf
)} , \label{S31}
\end{eqnarray}
where  $|{\cal E}(x)| = |{\rm det}{\cal E}(x)| $ is the volume of the lattice
element correspondent to the site $x$.

Here we introduce lattice couplings $\bar{\alpha}, \bar{\beta}, \bar{\delta},
\bar{\lambda}, \bar{\gamma}$ that differ from the original ones by the factors,
which are formally infinite and come from delta - functions in expressions
(\ref{RB}) and (\ref{TSING})  for torsion and curvature. We assume here, that a
certain regularization is made, which makes these factors finite. Our
supposition is that after the renormalization each physical quantity may be
expressed through physical couplings, which differ from the bare ones (both
lattice and continuum), and the infinity encountered here is absorbed into the
renormalization factors.

In our lattice model the translational connection $E^A_{\mu}(x)$ attached to
sites and the $SO(4)$ connection $U_{\mu}(x)$ attached to links are the
dynamical variables. The action of the model is expressed in a compact way
through these variables. It is easy to understand, that (\ref{trans}) is the
symmetry of the action. So, we have  lattice model with the direct
manifestation of Poincare gauge invariance.

As we have already noticed in the introduction, lattice realization of Poincare
quantum gravity has already been considered in several papers (see, for
example, \cite{Tomboulis, Magnea, Smolin, Manion, Kaku, Menotti} and references
therein). Now we are ready to compare our approach with the approaches of the
mentioned papers. First, in all of them the
 continuum models were transferred to lattice via application of naive
discretization procedure. This means, that certain variables attached to the
sites and/or links of the hypercubic lattice were considered, and it was set up
the correspondence between them and the Poincare group connection of the
continuum theory.  In \cite{Magnea,
 Kaku, Menotti} both vierbein and $SO(4)$ connection on the lattice were
attached to links. In \cite{Smolin} Poincare group was considered as a limiting
case of de Sitter group, and link variable belongs to $SO(5)$. In
\cite{Tomboulis, Manion} the $SO(4)$ connection was attached to links while the
vierbein was attached to sites of the lattice. The definition of the lattice
model in \cite{Manion} is in principle close to the resulting definition of our
model on the dual lattice. However, the resulting models are not identical.
First of all, in \cite{Manion} the action was of the Einstein - Hilbert form in
Palatini formulation, i.e. it contains only the first power in lattice
curvature. The term $-\bar{\gamma} m_p^2 \sum_{x\in sites}  {\bf R}(x)|{\cal
E}(x)|$ of (\ref{S31}) would coincide with the action of  \cite{Manion} if in
the latter the symmetrization over different directions at each site is
performed. In \cite{Tomboulis} the squared curvature action of general type was
considered. Both the link $O(4)$ connection and the site inverse vierbein were
considered in $4\times 4$ spinor representation. The final expression for the
lattice action is rather complicated and does not coincide with (\ref{S31}).

The crucial difference between the mentioned approaches and the approach of the
present paper is that we use the regular procedure and approximate
Riemann-Cartan manyfold via piecewize - linear Riemann-Cartan manyfolds. This
procedure gives us the discretization of the original continuum model, which is
manifestly invariant under general coordinate transformations. Contrary to
this, the constructions considered in \cite{Tomboulis, Magnea, Smolin, Manion,
Kaku, Menotti} violate general coordinate invariance and do not give gauge
invariant discretization of Poincare gravity.

\section{Measure over discretized geometries}

In order to define measure over dynamical variables in our lattice model we use
an analogy with  QCD, the lattice version of which is considered to work
perfectly. In QCD there are two kinds of fields:

1. Quarks and leptons. (The correspondent measure over Grassmann variables is
well defined and unique.)

2. The gauge field. The correspondent measure on the lattice is unique as it is
completely defined via symmetry properties: that is the local measure invariant
under gauge transformations.

In our case there are two fields: $SO(4)$ connection and the translational
connection. So, it is natural to use measures, which are invariant under
lattice realization of the gauge transformation. Our choice of measure is the
measure, which is simultaneously  invariant under lattice gauge transformations
and is local.

Each piecewise linear manyfold  described above is itself a Riemann - Cartan
space. Let the given discretization (with varying $E$ and $U$) be denoted as
$\cal M$. Then, we consider the set of correspondent independent variables
$\{E^A_{\mu}({\bf I}); U_{\bf IJ}\}$. Gauge transformation corresponds to the
shift of each lattice element by the vector $v^A ({\bf I})$ and its rotation
$\Theta_{\bf I} \in SO(4)$. This transformation acts as $\{E^A_{\mu}({\bf I});
U_{\bf IJ}\}\rightarrow \{\Theta_{\bf I}E_{\mu}({\bf I})+v({\bf I}));
\Theta_{\bf I}U_{\bf IJ}\Theta^T_{\bf J}\}$.

The locality of lattice measure means the following. The whole measure should
be represented as
\begin{equation}
D_{\cal M} (E;U) = \Pi_{\bf I} \Pi_{\mu}D E^A_{\mu}({\bf I}) \Pi_{\bf I, J}
DU_{\bf IJ},
\end{equation}
Here the product is over the sides of lattice elements
 and over the links that connect centers of lattice elements with their vertices.
 The measure over link matrices $U_{\bf IJ}$ is denoted by $DU_{\bf IJ}$. The
 measure over vectors $E^A_{\mu}({\bf I})$ is denoted by $D E^A_{\mu}({\bf
 I})$. We call the lattice measure local if inside each lattice element $DE^A_{\mu}$
for the given $\mu$ depends upon $E^A_{\mu}$ only,  and $DU_{\bf IJ}$ for the
given $\bf I, J$ depends upon $U_{\bf I J}$ only. It is obvious that this
requirement together with gauge invariance fixes the only choice of
$DE^A_{\mu}$ and $DU_{\nu}$: $D E^A_{\mu} = \Pi_{A,\mu} d E^A_{\mu}$, while
$DU$ is the invariant measure on $SO(4)$.

We must mention, that another locality principle can be formulated. Say, we may
thought that the measure is local if $DE^A_{\mu}$ may depend upon $E^A_{\nu}$
with $\nu \ne \mu$ but it may not depend upon $E^A_{\mu}$ from another lattice
element. Then gauge invariance does not fix measure precisely. However, we
choose the more strong requirement that was described above since it gives us
an opportunity to fix the only local gauge invariant measure.

\section{Metropolis algorithm.}

It is worth mentioning, that in real numerical calculations it would be useful
to express each $SO(4)$ link matrix (on the dual lattice) as the function of
the
 $SL(2,C)$ matrix. The correspondence is given by the conventional spinor
 representation of $SO(4)$ rotations. Then the invariant measure on $SL(2,C)$
  generates the invariant measure on $SO(4)$. It is much more easy to simulate the $SL(2,C)$
field than the $SO(4)$ field itself.

Metropolis algorithm for the simulation of our model can be described as
follows. At each step of the algorithm the given particular link and one of its
ends are considered. It is formed the proposition of the  link $SL(2,C)$ matrix
and the $4\times4$ matrix attached to the given endpoint of the
link\footnote{We may fix the gauge and choose $E^A_0 = 0 $. The correspondent
Faddeev-Popov determinant is constant.}. Then the terms of the action
(\ref{S31}), which contain torsion and correspond to the neighbors of the given
point are calculated. Next, we calculate the terms of the action, which contain
curvature and correspond to the points of the "butterfly" correspondent to the
given link. The "butterfly" is the figure that consists of all plaquettes with
 the given link as one of their sides. Let the sum of these terms be denoted
as $S^{\rm new}$. Then let the same sum that was calculated using the old
values of the proposed  variables, be denoted as $S^{\rm old}$. The proposition
is accepted if $S^{\rm new}<S^{\rm old}$. Otherwise it is accepted with the
probability ${\rm exp}\,(S^{\rm old}-S^{\rm new})$. Thereafter we choose
another link and one of its ends, and the procedure is repeated.

In order to accelerate the numerical simulation one can save in the computer
memory the values of all plaquette variables $\Omega_{\mu \nu}$, the values of
all lattice derivatives $\Delta_{\mu} {\cal E}_{\nu}$, and the values of the
inverse matrices $\{{\cal E}^A_{\mu}\}^{-1}$. This would lead to an essential
economy of CPU time but requires an additional memory size.

\section{Conclusions}

In this paper we construct the hypercubic discretization of Poincare quantum
gravity. The main achievement of our construction, which differs it from the
other discretizations of Poincare gravity, is that this construction is
manyfestly gauge invariant (with respect to the group of general coordinate
transformations). Namely, we approximate continuum Riemann-Cartan manyfold by
piecewise-linear Riemann-Cartan manyfolds. The resulting lattice model deals
with the invariant geometrical properties of these piecewise-linear manyfolds.
This construction is analogous to Regge discretization of Riemannian manyfolds
and is its direct generalization.

We consider piecewise - linear manyfolds composed of flat pieces of hypercubic
form. So, the geometry is defined by the forms and sizes of these flat pieces
together with the $SO(4)$ rotation matrices attached to their sides. Actually
these variables are expressed through the translational connection (which is
constant inside each flat piece) and the $SO(4)$ connection (which is singular
and is attached to the sides of the hypercubes). Therefore, these translational
and rotational connections are the dynamical variables of the discretized
model.

The piecewise-linear manyfold has singular torsion and curvature, which are
calculated directly and expressed through the sizes of the lattice elements and
the rotation matrices.

In addition, we calculated  the squared curvature action of rather general form
on the piecewise-linear manyfold. It is also expressed through the dynamical
variables of the discretization.

Next, we point out that the lattice model may be considered as a lattice
Poincare gauge theory. Namely, shift and rotation of each piece of the given
piecewise-linear manyfold with respect to the local map cause the
transformation of the chosen dynamical variables. The lattice action is
invariant under this transformation.

We construct local measure over lattice dynamical variables that is invariant
under the lattice Poincare transformations.

Finally, we obtain gauge invariant lattice realization of Poincare quantum
gravity, in which the translational connection ($4\times4$ matrix) attached to
the sites of the dual lattice and the $SO(4)$ matrix (or, equivalently,
$SL(2,C)$ matrix) attached to links of the dual lattice play the role of the
dynamical variables. The action of the lattice model contains only the first
lattice derivatives of the dynamical variables and has a compact and rather
simple form. Thus the constructed lattice model is expected to be useful for
the numerical simulations. The correspondent Metropolis algorithm is briefly
described.

This work was partly supported by RFBR grants 06-02-16309, 05-02-16306, and
04-02-16079, by Federal Program of the Russian Ministry of Industry, Science
and Technology No 40.052.1.1.1112.

\section{Appendix}
Here we represent the expressions for curvature and torsion on the
piecewise-linear manyfold.

Connection is singular on the sides of lattice elements. $SO(4)$ curvature is
concentrated on the bones\footnote{On the hypercubic lattice we refer to the
plaquettes as to the bones in order to make an analogy with the simplicial
Regge calculus.}. We choose the following integral equation as a definition of
$SO(4)$ curvature.
\begin{eqnarray}
&& 2 \int_{y\in{\Sigma}}\Omega(z,y) R_{\mu\nu}(y)\Omega^+(z,y)dy^{\mu}\wedge
dy^{\nu} = P {\rm exp} (\int^z_{z\in \partial \Sigma} \omega_{\mu} dx^{\mu})
\nonumber\\&& - [P {\rm exp} (\int^z_{z\in \partial \Sigma} \omega_{\mu}
dx^{\mu})]^+ \, {\rm at} \, |\Sigma|\rightarrow 0
\end{eqnarray}

Here $\omega_{\mu}$ is $SO(4)$ connection\footnote{$U_{IJ}=P {\rm exp} (\int
\omega_{\mu} dx^{\mu})$, where integral is over the path of minimal length,
which connects centers of lattice elements $I$ and $J$.}. $\Sigma$ is a small
surface, that crosses the given bone, and $ |\Sigma|$ is its area. $\partial
\Sigma$ is the boundary of $\Sigma$. Its orientation corresponds to orientation
of $\Sigma$. $\Omega(z,y) = P {\rm exp} (\int^y_z \omega_{\mu} dx^{\mu})$ is
the parallel transporter along the path that connects a fixed point on
$\partial \Sigma$ with the point $y$. We choose this path in such a way, that
it is winding around the given bone in the same direction as $\partial \Sigma$
and has minimal length. The integral in the right hand side is over the path
$\partial \Sigma$, which begins and ends at the point $z$.

It is worth mentioning, that the given definition does not contradict with the
conventional one in case of smooth connection. And it gives us the possibility
to calculate curvature in the case of the constructed singular piecewise -
linear manyfold.

 Let us fix the
given lattice element. Inside it lattice curvature is equal to
\begin{equation}
R_{\mu\nu B}^A(y) = \sum_{b\in bones}\int_{x\in{b}}\epsilon_{\mu \nu \rho
\sigma}dx^{\rho}\wedge dx^{\sigma}\delta^{(4)}(y-x)
\frac{[\Pi_{i}U^b_{I_iI_{i+1}}-
(\Pi_{i}U^b_{I_iI_{i+1}})^+]^A_B}{2D!},\label{RB1}
\end{equation}
Here the sum is over the bones that belong to the given lattice element. The
integral is over the surface of the bone. The product $\Pi_{i}U^b_{I_iI_{i+1}}$
of the rotation matrices is along the closed path around the given bone $b$,
which consists of links that connect centers of the lattice elements. Here we
imply, that this closed path begins within the given lattice element and has
the minimal lattice length.

Now let us calculate torsion, which is concentrated on the sides of lattice
elements. The torsion field $T_{\mu \nu}^A$ is defined by the integral equation
\begin{equation}
 \int_{y\in{\Sigma}}\Omega^A_B(z,y)
T^B_{\mu\nu}(y)dy^{\mu}\wedge dy^{\nu} = \int_{\partial \Sigma}\Omega^A_B(z,y)
b^B_{\mu}(y)dy^{\mu}
\end{equation}
Here $b^A_{\mu}(x)$ is the field of inverse vierbein, which is expressed
through our variables $E^A_{\mu}$ inside each lattice element if the given
parametrization of the lattice element is chosen.

This equation is satisfied with the following expression (which is valid within
the lattice element $\bf I$):
\begin{eqnarray}
T^A_{\mu\nu}(y) & = & \sum_{s\in sides} [\int^{\bf J^s}_{x\in{s}}\frac{[U_{\bf
 IJ^s}]^A_B b^B_{[\mu}(x)\epsilon_{\nu] \tau \rho \sigma}}{D!}dx^{\tau}\wedge dx^{\rho}\wedge
dx^{\sigma}\delta^{(4)}(y-x)\nonumber\\
&& - \int^{\bf I}_{x\in{s}}\frac{b^A_{[\mu}(x)\epsilon_{\nu ]\tau \rho
\sigma}}{D!} dx^{\tau}\wedge dx^{\rho}\wedge
dx^{\sigma}\delta^{(4)}(y-x)]\label{TT}
\end{eqnarray}
Here the first integral in the sum is over the given side $s$ seing from the
neighbor lattice element $\bf J^s$ (the side $s$ is common for $\bf I$ and $\bf
J^s$). (We imply that  in (\ref{TT}) the given lattice element and all its
neighbors have the common parametrization.)

Let us define inside each lattice element the following variables: ${\cal
E}^A_{\mu} = E^A_{\mu} - E^A_0, \mu = 1,2,3,4$. Also we denote by ${\cal
E}^{\mu}_A$ elements of the inverse matrix ${\cal E}^{-1}$ In tetrad components
we have:
\begin{equation}
R_{C F B}^A(y) = {\cal E}^{\mu}_{C} {\cal E}^{\nu}_{F}\sum_{b\in
bones}\int_{x\in{b}}\epsilon_{\mu \nu \rho \sigma}dx^{\rho}\wedge
dx^{\sigma}\delta^{(4)}(y-x) \frac{[\Pi_{i}U^b_{I_iI_{i+1}}-
(\Pi_{i}U^b_{I_iI_{i+1}})^+]^A_B}{2D!},\label{RB}
\end{equation}

Torsion is expressed as
\begin{eqnarray}
T^A_{C F}(y) & = & \frac{{\cal E}^{\mu}_{{\bf I}C} {\cal E}^{\nu}_{{\bf
I}F}}{D!} \sum_{s\in sides} \int_{x\in{s}} dx^{\tau}\wedge dx^{\rho}\wedge
dx^{\sigma}\delta^{(4)}(y-x) \nonumber\\&&([U_{\bf
 IJ^s}]^A_B {\cal E}^B_{{\bf J^s}[\mu}\epsilon_{\nu] \tau \rho \sigma} - {\cal E}^A_{{\bf I}
[\mu}\epsilon_{\nu ]\tau \rho \sigma})\label{TSING}
\end{eqnarray}
Here $ {\cal E}^B_{{\bf I}\mu}$ is calculated inside the given lattice element
$\bf I$ while $ {\cal E}^B_{{\bf J^s}\mu}$ is calculated within its neighbor
$\bf J^s$.

\end{document}